# Dynamics, Computation, and the "Edge of Chaos": A Re-Examination

Melanie Mitchell[1], James P. Crutchfield[2], and Peter T. Hraber[1]

**Abstract**

In this paper we review previous work and present new work concerning the relationship between dynamical systems theory and computation. In particular, we review work by Langton [22] and Packard [29] on the relationship between dynamical behavior and computational capability in cellular automata (CA). We present results from an experiment similar to the one described in [29], that was cited there as evidence for the hypothesis that rules capable of performing complex computations are most likely to be found at a phase transition between ordered and chaotic behavioral regimes for CA (the "edge of chaos"). Our experiment produced very different results from the original experiment, and we suggest that the interpretation of the original results is not correct. We conclude by discussing general issues related to dynamics, computation, and the "edge of chaos" in cellular automata.

## 1. Introduction

A central goal of the sciences of complex systems is to understand the laws and mechanisms by which complicated, coherent global behavior can emerge from the collective activities of relatively simple, locally interacting components. Given the diversity of systems falling into this broad class, the discovery of any commonalities or "universal" laws underlying such systems will require very general theoretical frameworks. Two such frameworks are dynamical systems theory and the theory of computation. These have independently provided powerful tools for understanding and describing common properties of a wide range of complex systems.

Dynamical systems theory has developed as one of the main alternatives to analytic, closed-form, exact solutions of complex systems. Typically, a system is considered to be "solved" when one can write down a finite set of finite expressions that can be used to predict the state of the system at time $t$, given the state of the system at some initial time $t_0$. Using existing mathematical methods, such solutions are generally not possible for most complex systems of interest. The central contribution of dynamical systems theory to modern science is that exact solutions are not necessary for understanding and analyzing a nonlinear process. Instead of deriving exact single solutions, the emphasis of dynamical systems theory is on describing the geometrical and topological structure of ensembles of solutions. In other words, dynamical systems theory gives a geometric view of a process's structural elements, such as attractors, basins, and separatrices. It is thus distinguished from a purely probabilistic approach such as statistical mechanics, in which geometric structures

[1]Santa Fe Institute, 1660 Old Pecos Trail, Suite A, Santa Fe, New Mexico, U.S.A. 87501.
Email: mm@santafe.edu, pth@santafe.edu
[2]Physics Department, University of California, Berkeley, CA, U.S.A. 94720.
Email: chaos@gojira.berkeley.edu

are not considered. Dynamical systems theory also addresses the question of what structures are generic; that is, what behavior types are typical across the spectrum of complex systems

In contrast to focusing on how geometric structures are constrained in a state space, computation theory focuses on how basic information processing elements—storage, logical gates, stacks, queues, production rules, and the like—can be combined to effect a given information-processing task. As such, computation theory is a theory of organization and the functionality supported by organization. When adapted to analyze complex systems, it provides a framework for describing behaviors as computations of varying structure. For example, if the global mapping from initial to final states is considered as a computation, then the question is, what function is being computed by the global dynamics? Another range of examples concern limitations imposed by the equations of motion on information processing: can a given complex system be designed to emulate a universal Turing Machine? In contrast to this sort of engineering question, one is also interested in the intrinsic computational capability of a given complex system; that is: what information-processing structures are intrinsic in its behavior? [9, 17]

Dynamical systems theory and computation theory have almost always been applied independently, but there have been some efforts to understand the relationship between the two—that is, the relationship between a system's ability for information processing and other measures of the system's dynamical behavior.

**Relationships Between Dynamical Systems Theory and Computation Theory**

Computation theory developed from the attempt to understand information-processing aspects of systems. A colloquial definition of "information processing" might be "the transformation of a given input to a desired output", but in order to apply the notion of information processing to complex systems and to relate it to dynamical systems theory, the notion must be enriched to include the *production* of information as well as its storage, transmission, and logical manipulation. In addition, the engineering-based notion of "desired output" is not necessarily appropriate in this context; the focus here is often on the intrinsic information-processing capabilities of a dynamical system not subject to a particular computational goal.

Beginning with Kolmogorov's and Sinai's adaptation of Shannon's communication theory to mechanics in the late 1950's [20, 33], there has been a continuing effort to relate a nonlinear system's information processing capability and its temporal behavior. One result is that a deterministic chaotic system can be viewed as a generator of information [32]. Another is that the complexity of predicting a chaotic system's behavior grows exponentially with time [7]. The descriptive complexity here, called the Kolmogorov-Chaitin complexity [4, 21], uses a universal Turing machine as the deterministic prediction machine. The relationship between the difficulty of prediction and dynamical randomness is simply summarized by the statement that the growth rate of the descriptive complexity is equal to the information production rate [3]. These results give a view of deterministic chaos that emphasizes the production of randomness and the resulting unpredictability. They are probably the earliest connections between dynamics and computation.

The question of what structures underlie information production in mechanical systems has received attention only more recently. The first and crudest property considered is the amount of memory a system employs in producing apparent randomness [8, 15]. The idea is that an ideal random process uses no memory to produce its information—it simply flips a coin as needed. Similarly, a simple periodic process requires memory only in proportion to the length of the pattern it repeats. Within the memory-capacity view of dynamics, both these types of processes are simple—more precisely, they are simple to describe statistically. Between these extremes, though, lie the highly structured, complex processes that use both randomness and pattern storage to produce their behavior. Such processes are more complex to describe statistically than are ideal random or simple periodic processes. The trade-off between structure and randomness is common to much of science. The notion of statistical complexity [9] was introduced to measure this trade-off.

Computation theory is concerned with more than information and its production and storage. These elements are taken as given and, instead, the focus is on how their combinations yield more or less computational power. Understandably, there is a central dichotomy between machines with finite and infinite memory. On a finer scale, distinctions can be drawn among ways in which infinite memory is organized—e.g., as a stack, a queue, or a parallel array. Given such considerations, the question of the intrinsic computational structure in a dynamical system becomes substantially more demanding than the initial emphasis on gross measures of information storage and production.

Several connections in this vein have been made recently. In the realm of continuous-state dynamical systems, Crutchfield and Young looked at the relationship between the dynamics and computational structure of discrete time series generated by the logistic map at different parameter settings [9, 10]. They found that at the onset of chaos there is an abrupt jump in computational class of the time series, as measured by the formal language class required to describe the time series. In concert with Feigenbaum's renormalization group analysis of the onset of chaos [12], this result demonstrated that a dynamical system's computational capability—in terms of the richness of behavior it produces—is qualitatively increased at a phase transition.

Rather than considering intrinsic computational structure, a number of "engineering" suggestions have been made that there exist physically plausible dynamical systems implementing Turing machines [2, 25, 26]. These studies provided explicit constructions for several types of dynamical systems. At this point, it is unclear whether the resulting computational systems are generic—i.e., likely to be constructable in other dynamical systems—and whether they are robust and reliable in information processing. In any case, it is clear that much work has been done to address a range of issues that relate continuous-state dynamics and computation. Many of the basic issues are now clear and there is a firm foundation for future work.

### Dynamics and Computation in Cellular Automata

There has also been a good deal of study of dynamics and computation in discrete spatial systems called cellular automata (CA). In many ways, CA are more natural candidates for this study than continuous-state dynamical systems since they are completely discrete in space, in time, and in local state. There is no need to develop a theory of computation with real numbers. Unfortunately, something is lost in going to a completely discrete system. The analysis of CA behavior in conventional dynamical systems terms is problematic for just this reason. Defining the analogs of "sensitive dependence on initial conditions", "the production of information", "chaos", "instability", "attractor", "smooth variation of a parameter", "bifurcation", the "onset of chaos", and other basic elements of dynamical systems theory requires a good deal of care. Nonetheless, Wolfram introduced a dynamical classification of CA behavior closely allied to that of dynamical systems theory. He speculated that one of his four classes supports universal computation [35]. It is only recently, however, that CA behavior has been directly related to the basic elements of qualitative dynamics—the attractor-basin portrait [17]. This has lead to a reevaluation of CA behavior classification and, in particular, to a redefinition of the chaos and complexity apparent in the spatial patterns that CA generate [6].

Subsequent to Wolfram's work, Langton studied the relationship between the "average" dynamical behavior of cellular automata and a particular statistic ($\lambda$) of a CA rule table [22]. He then hypothesized that "computationally capable" CA, and in particular, CA capable of universal computation, will have "critical" $\lambda$ values corresponding to a phase transition between ordered and chaotic behavior. Packard experimentally tested this hypothesis by using a genetic algorithm (GA) to evolve CA to perform a particular complex computation [29]. He interpreted the results as showing that the GA tends to select rules close to "critical" $\lambda$ regions—i.e., the "edge of chaos".

We now turn our discussion more specifically to issues related to $\lambda$, dynamical-behavior classes, and computation in CA. We then present experimental results and a theoretical discussion that suggest the interpretation given of the results in [29] is not correct. Our experiments, however, show some interesting phenomena with respect to the GA evolution of CA, which we summarize here. A longer, more detailed description of our experiments and results is given in [24].

## 2. Cellular Automata and the "Edge of Chaos"

Cellular automata are one of the simplest frameworks in which issues related to complex systems, dynamics, and computation can be studied. CA have been used extensively as models of physical processes and as computational devices [11, 16, 30, 34, 36]. In its simplest form, a CA consists of a spatial lattice of *cells*, each of which, at time $t$, can be in one of $k$ states. We denote the lattice size (i.e., number of cells) as $N$. A CA has a single fixed rule used to update each cell; this rule maps from the states in a neighborhood of cells—e.g., the states of a cell and its nearest neighbors—to a single state, which becomes the updated value for the cell in question. The lattice starts out with some initial configuration of cell states and, at each time step, the states of all cells in the lattice are synchronously updated. We use the term "state" to refer to the value of a single cell —e.g., 0 or 1—and the term "configuration" to mean the pattern of states over the entire lattice.

In this paper we restrict our discussion to one-dimensional CA with $k = 2$. In a one-dimensional CA, the neighborhood of a cell includes the cell itself and some number $r$ of neighbors on either side of the cell. All of the simulations described here are of CA with spatially periodic boundary conditions (i.e., the one-dimensional lattice is viewed as a circle, with the right neighbor of the rightmost cell being the leftmost cell, and vice versa).

The equations of motion $\phi$ for a CA are often expressed in the form of a *rule table*. This is a lookup table listing each of the neighborhood patterns and the state to which the central cell in that neighborhood is mapped. For example, the following is one possible rule table for a one-dimensional CA with $k = 2, r = 1$. Each possible neighborhood $\eta$ is given along with the "output bit" $s = \phi(\eta)$ to which the central cell is updated.

| $\eta$ | 000 | 001 | 010 | 011 | 100 | 101 | 110 | 111 |
|---|---|---|---|---|---|---|---|---|
| $s$ | 0 | 0 | 0 | 1 | 0 | 1 | 1 | 1 |

In words, this rule says that for each neighborhood of three adjacent cells, the new state is decided by a majority vote among the three cells.

The notion of "computation" in CA can have several possible meanings [24], but the most common meaning is that the CA performs some "useful" computational task. Here, the rule is interpreted as the "program", the initial configuration is interpreted as the "input", and the CA runs for some specified number of time steps or until it reaches some "goal" pattern—possibly a fixed-point pattern. The final pattern is interpreted as the "output". An example of this is using CA to perform image-processing tasks [31].

Packard [29] discussed a particular $k = 2, r = 3$ rule, invented by Gacs, Kurdyumov, and Levin (GKL) [13] as part of their studies of reliable computation in CA. The GKL rule was not invented for any particular classification purpose, but it does have the property that, under the rule, most initial configurations with less than half 1's are eventually transformed to a configuration of all 0's, and most initial configurations with more than half 1's are transformed to a configuration of all 1's. The rule thus approximately computes whether the density of 1's in the initial configuration (which we denote as $\rho$) is above the threshold $\rho_c = 1/2$. When initial configurations are close to $\rho = 1/2$, the rule makes a significant number of classification errors [24].

Packard was inspired by the GKL rule to use a GA to *evolve* a rule table to perform this "$\rho_c = 1/2$" task. If $\rho < 1/2$ then the CA should relax to a configuration of all 0's; otherwise it should relax to a configuration of all 1's. This task can be considered to be a "complex" computation for a $k = 2, r = 3$ CA since the minimal amount of memory it requires increases with $N$; in other words, the required computation is spatially global and corresponds to the recognition of a non-regular language.[3] The global nature of the computation means that information must be transmitted over significant space-time distances (on the order of $N$) and this requires the cooperation of many local neighborhood operations [24].

In dynamical terms, complex computation in a small-radius, binary-state CA requires significantly long transients and space-time correlation lengths. Langton hypothesized that such effects are most likely to be seen in a certain region of CA rule space as parameterized by $\lambda$ [22]. For binary-state CA, $\lambda$ is simply the fraction of 1's in the output bits of the rule table. For CA with $k > 2$, $\lambda$ is defined as the fraction of "non-quiescent" states in rule table, where one state is arbitrarily chosen to be "quiescent", and all states obey a "strong quiescence" requirement [22]. Langton performed a number of Monte Carlo samples of two-dimensional CA, starting with $\lambda = 0$ and gradually increasing $\lambda$ to $1 - 1/k$ (i.e., the most homogeneous to the most heterogeneous rule tables). Langton used various statistics such as single-site entropy, two-site mutual information, and transient length to classify CA "average" behavior at each $\lambda$ value. The notion of "average behavior" was intended to capture the most likely behavior observed with a randomly chosen initial configuration for CA randomly selected in a fixed-$\lambda$ subspace. These studies revealed some correlation between the various statistics and $\lambda$. The correlation is quite good for very low and very high $\lambda$ values. However, for intermediate $\lambda$ values in finite-state CA, there is a large degree of variation in behavior.

---

[3]See [19] for an introduction to formal-language classes in computation theory.

Langton claimed on the basis of these statistics that as $\lambda$ is incremented from 0 to $[1 - 1/k]$, the average behavior of CA undergoes a "phase transition" from ordered (fixed point or limit cycle after some short transient period) to chaotic (apparently unpredictable after some short transient period). As $\lambda$ reaches a "critical value" $\lambda_c$, the claim is that rules tend to have longer and longer transient phases. Additionally, Langton claimed that CA close to $\lambda_c$ tend to exhibit long-lived, "complex"—non-periodic, but non-random—patterns. Langton proposed that the $\lambda_c$ regime roughly corresponds to Wolfram's Class 4 CA [35], and hypothesized that CA able to perform complex computations will most likely be found in this regime.

Analysis based on $\lambda$ is one possible first step in understanding the structure of CA rule space and the relationship between dynamics and computation in CA. However, the claims summarized above rest on a number of problematic assumptions. One assumption is that in the global view of CA space, CA rule tables themselves are the appropriate loci of dynamical behavior. This is in stark contrast with the state space and the attractor-basin portrait approach of dynamical systems theory. The latter approach acknowledges the fact that behaviors in state space cannot be adequately parameterized by any function of the equations of motion, such as $\lambda$. Another assumption is that the underlying statistics being averaged (e.g., single-site entropy) converge. But many processes are known for which averages do not converge. Perhaps most problematic is the assumption that the selected statistics are uniquely associated with mechanisms that support useful computation.

Packard empirically determined rough values of $\lambda_c$ for one-dimensional $k = 2, r = 3$ CA by looking at the *difference-pattern spreading rate* $\gamma$ as a function of $\lambda$ [29]. The spreading rate $\gamma$ is a measure of unpredictability in spatio-temporal patterns and so is one possible measure of chaotic behavior [27, 35]. It is analogous to, but not the same as, the Lyapunov exponent for continuous-state dynamical systems. In the case of CA it indicates the average propagation speed of information through space-time, though not the production rate of local information. At each $\lambda$ a large number of rules was sampled and for each CA $\gamma$ was estimated. The average $\gamma$ over the selected CA was taken as the average spreading rate at the given $\lambda$. The results are reproduced in Figure 1(a). As can be seen, at low and high $\lambda$'s, $\gamma$ vanishes, indicating fixed-point or short-period behavior; at intermediate $\lambda$ it is maximal, indicating chaotic behavior; and in the transition or $\lambda_c$ regions—centered about $\lambda \approx 0.25$ and $\lambda \approx 0.80$—it rises or falls gradually. While not shown in Figure 1(a), for most $\lambda$ values $\gamma$'s variance, like that of the statistics used by Langton, is high.

## 3. The Original Experiment

The empirical CA studies recounted above addressed only the relationship between $\lambda$ and the dynamical behavior of CA as revealed by several statistics. Those studies did not correlate $\lambda$ or behavior with an independent measure of computation. Packard [29] addressed this issue by using a genetic algorithm (GA) [14, 18] to evolve CA rules to perform a particular computation. This experiment was meant to test two hypotheses: (1) CA rules able to perform complex computations are most likely to be found near $\lambda_c$ values; and (2) when CA rules are evolved to perform a complex computation, evolution will tend to select rules near $\lambda_c$ values.

Packard's experiment consisted of evolving binary-state one-dimensional CA with $r = 3$. The "complex computation" is the $\rho = 1/2$ task described above. A form of the genetic algorithm was applied to a population of rules represented as bit strings. To calculate the fitness of a string, the string was interpreted as the output bits of a rule table, and the resulting CA was run on a number of randomly chosen initial conditions. The fitness was a measure of the average classification performance of the CA over the initial conditions.

The result from this experiment are displayed in Figure 1(b). The histogram displays the observed frequency of rules in the GA population as a function of $\lambda$, with rules merged from a number of different runs with identical parameters but with different random number seeds. In the initial generation the rules were uniformly distributed over $\lambda$ values. The graph (b) gives the final generation—in this case, after the GA has run for 100 generations. The rules cluster close to the two $\lambda_c$ regions, as can be seen by comparison with the difference-pattern spreading rate plot (a). Note that each individual run produced rules at one or the other peak in graph (b), so when the runs were merged together, both peaks appear [28]. Packard interpreted these results as evidence for the hypothesis that, when an ability for complex computation is required, evolution tends to select rules near the transition to chaos. He argues, like Langton, that this result intuitively makes sense because "rules near the transition to chaos have the capability to selectively communicate information with complex structures in space-time, thus enabling computation." [29].

## 4. Our Experiment

We performed an experiment similar to Packard's. The CA rules in the population are represented as bit strings, each encoding the output bits of a rule table for $(k, r) = (2, 3)$. Thus, the length of each string is $128 = 2^{2r+1}$.

For a single run, the GA we used generated a random initial population of 100 rules (bit strings) with $\lambda$ values uniformly distributed over [0,1]. Then it calculated the fitness of each rule in the population by a method to be described below. The population was then ranked by fitness and the 50 rules with lowest fitness were discarded. The 50 rules with highest fitness were copied directly into the next generation. To fill out the population 50 new rules were generated from pairs of parents selected at random from the current generation. Each pair of parents underwent a single-point crossover whose location was selected with uniform probability over the string. The resulting offspring were mutated at a number of sites chosen from a Poisson distribution with a mean of 3.8.

The fitness of a rule $R$ is calculated as follows. $R$ is run on 300 randomly chosen initial configurations on a lattice with $N = 149$. A new set of initial configurations is chosen each generation, and all rules in that generation are tested on it. The 300 initial configurations are uniformly distributed over densities in [0,1], with exactly half having $\rho < 1/2$ and exactly half having $\rho > 1/2$. $R$ is run on each initial configuration for approximately 320 iterations; the actual number is chosen probabilistically to avoid overfitting. 320 iterations is the measured maximum amount of time for the GKL CA to reach an invariant pattern over a large number of initial configurations on lattice size 149.

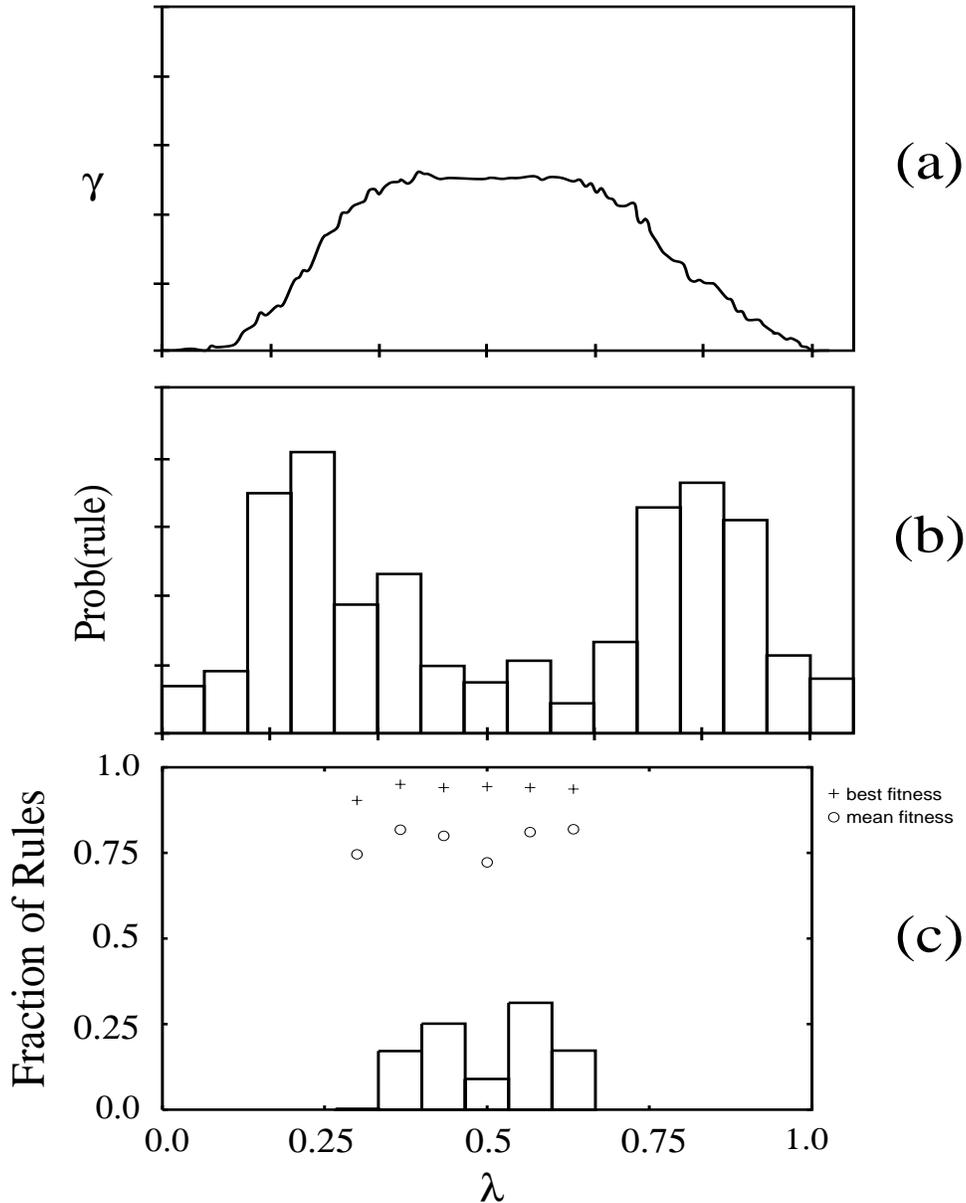

Figure 1: (a): The average difference-pattern spreading rate $\gamma$ of a large number of randomly chosen $k = 2, r = 3$ CA, as a function of $\lambda$.
(b): Results from the original experiment on GA evolution of CA for the $\rho_c = 1/2$ classification task. The histogram plots the frequencies of rules merged from the final generations (generation 100) of a number of runs. These populations evolved from initial populations uniformly distributed in $\lambda$. The histogram consists of 16 bins of width 0.0667. The bin above $\lambda = 1.0$ contains just those rules with $\lambda = 1.0$. Graphs (a) and (b) are adapted from [29], with the author's permission. No vertical scale was provided there.
(c): Results from our experiment. The histogram plots the frequencies of rules merged from the final generations (generation 100) of 30 runs. These populations evolved from initial populations uniformly distributed in $\lambda$. Following [29] the $\lambda$-axis is divided into 15 bins of length 0.0667 each. The rules with $\lambda = 1.0$ are included in the rightmost bin. The best (cross) and mean (circle) fitnesses are plotted for each bin. (The $y$-axis interval for fitnesses is also [0,1]).

$R$'s score on a given initial configuration is the fraction of "correct" bits in the final configuration. For example, if the initial configuration has $\rho > 1/2$ then $R$'s score is the fraction of 1's in the final configuration. Thus, $R$ gets partial credit for getting some of the bits correct. A rule generating random strings would therefore get a score of 0.5. $R$'s fitness is then its average score over all 300 initial configurations. For more details and for justifications for these parameters, see [24].

The results of our experiment are given in Figure 1(c). This histogram displays the observed frequency of rules in the population at generation 100 as a function of $\lambda$, merged from 30 different runs with identical parameters but different random number seeds. The initial populations were each uniformly distributed over $\lambda$. The best and mean fitnesses of rules in each bin are also displayed.

There are a number of striking differences between Figures 1(b) and 1(c):

- In Figure 1(b), most of the rules in the final generations cluster in the $\lambda_c$ regions defined by Figure 1(a). In particular, in Figure 1(b), approximately 66% of the mass of the distribution is in bins 3–5 and 12–14 combined (where bins are numbered 1–16 left to right). In Figure 1(c) these bins contain only 0.002% of the mass of the distribution (there are no rules in bins 3, 4, 12, 13, or 14, and there are only 5 rules in bin 5 out of a total of 3000 rules represented in the histogram).

- In Figure 1(b) there are rules in every bin. In Figure 1(c) there are rules only in the central six bins.

- In both histograms there are two peaks surrounding a central dip. As in the original experiment, in our experiment each individual run produced rules at one or the other peak, so when the runs were merged together, both peaks appear. In Figure 1(b), however, the two peaks are located roughly at bins 4 and 13 and thus are centered around $\lambda = 0.23$ and $\lambda = 0.83$, respectively. In Figure 1(c) the peaks are located roughly at bins 7 and 9 and thus are centered around $\lambda = 0.43$ and $\lambda = 0.57$, respectively. The ratio of the two peak spreads is thus approximately 4:1.

- In Figure 1(b), the two highest bins are roughly five times as high as the central bin whereas in Figure 1(c) the two highest bins are roughly three times as high as the central bin.

Figure 1(c) also gives an important calibration: the best and mean fitness of rules in each bin. The best fitnesses are all between 0.93 and 0.95, except the leftmost bin which has a best fitness of 0.90. Under this fitness function the GKL rule has fitness $\approx 0.98$ on all lattice sizes; the GA never found a rule with fitness above 0.95 on lattice size 149, and the measured fitness of the best evolved rules was much worse on larger lattice sizes [24]. The fitnesses of the rules in Figure 1(b) were not given in [29], though none of those rules achieved the fitness of the GKL rule [28].

## 5. Discussion of Experimental Results

### Why Do the Rules Cluster Close to $\lambda = 1/2$?

What accounts for these differences between Figures 1(b) and 1(c)? In particular, why did the evolved rules in our experiment tend to cluster close to $\lambda = 1/2$ rather than the two $\lambda_c$ regions?

There are two reasons (discussed in detail below): (1) Good performance on the $\rho_c = 1/2$ task *requires* rules with $\lambda$ close to $1/2$; and (2) The GA operators of crossover and mutation intrinsically push any population close to $\lambda = 1/2$.

It can be shown that correct or nearly correct performance on the $\rho_c = 1/2$ task requires rules close to $\lambda = 1/2$. Intuitively, this is because the task is symmetric with respect to the exchange of 1's and 0's. Suppose, for example, a rule that carries out the $\rho_c = 1/2$ task has $\lambda < 1/2$. This implies that there are more neighborhoods in the rule table that map to output bit 0 than to output bit 1. This, in turn, means that there will be *some* initial configurations with $\rho > \rho_c$ on which the action of the rule will *decrease* the number of 1's. And this is the opposite of the desired action. However, if the rule acts to *decrease* the number of 1's on an initial configuration with $\rho > \rho_c$, it risks producing an intermediate configuration with $\rho < \rho_c$, which then would lead (under the original assumption that the rule carries out the task correctly) to a fixed point of all 0's, misclassifying the initial configuration. A similar argument holds in the other direction if the rule's $\lambda$ value is greater than $1/2$. This informal argument shows that a rule with $\lambda \neq 1/2$ will misclassify certain initial configurations. Generally, the further away the rule is from $\lambda = 1/2$, the more such initial configurations there will be. Such rules may perform fairly well, classifying many initial configurations correctly or partially correctly. However, we expect any rule that performs reasonably well on this task—in the sense of being close to the GKL rule's 0.98 average fitness across lattice sizes—to have a $\lambda$ value close to $1/2$. This is one force pushing the GA population to $\lambda = 1/2$. We note that, not surprisingly, the GKL rule has $\lambda = 1/2$.

This analysis points to a problem with using this task as an evolutionary goal in order to test the hypothesis relating evolution, computation, and $\lambda_c$ rules. As was shown in Figure 1(a), for $k = 2, r = 3$ CA the $\lambda_c$ values occur at roughly 0.25 and 0.80. But for the $\rho$-classification tasks, the range of $\lambda$ values required for good performance is simply a function of the task and, specifically, of $\rho_c$. For example, the underlying 0-1 exchange symmetry of the $\rho_c = 1/2$ task implies that if a CA exists to do the task at an acceptable performance level, then it has $\lambda \approx 1/2$. Even though this basic point does not directly invalidate the hypothesis concerning evolution to $\lambda_c$ regions or claims about $\lambda$'s correlation with *average* behavior, it presents problems with using $\rho$-classification tasks as a way to gain evidence about a generic relation between $\lambda$ and computational capability. In our view, though, useful general hypotheses about evolution and computation should apply at least to computational tasks such as density classification.

A second force pushing rules to cluster close to $\lambda = 1/2$ is a "combinatorial drift" force, by which the random actions of crossover and mutation, apart from any selection force, tend to push the population towards $\lambda = 1/2$. The results of experiments measuring the relative effects of this force and the selection force in our experiment are given in [24].

Our experimental results, along with the theoretical argument that the most successful rules for this task should have $\lambda$ close to $1/2$, lead us to conclude that it is not correct to interpret Figure 1(b) as evidence for the hypothesis that CA able to perform complex computations will most likely be found close to $\lambda_c$. This is an important conclusion, since [29] is the only published experimental study directly linking $\lambda$ with computational ability in CA.

In appreciating this, one must keep in mind that it has been known for some time that some CA, e.g. the Game of Life CA, are capable in principle of universal computation [1]. The Game of Life has $\lambda \approx \lambda_c$. Langton [23] demonstrated that another two-dimensional CA with $\lambda \approx \lambda_c$ is capable in principle of universal computation, using a construction similar to the proof of computation universality for the game of Life. However, as Langton points out, these particular constructions do not establish any necessary correlation between $\lambda_c$ and the ability for complex, or even universal, computation.

As far as the GA results are concerned, we do not know what accounted for the differences between our results and those obtained in the original experiment. We speculate that the differences are due to additional mechanisms in the GA used in the original experiment that were not reported in [29]. For example, the original experiment included a number of additional sources of randomness, such as the regular injection of new random rules at various $\lambda$ values and a much higher mutation rate than that in our experiment [28]. These sources of randomness may have slowed the GA's search for high-fitness rules and prevented it from converging on rules close to $\lambda = 1/2$. Our experimental results and theoretical analysis give strong reason to believe that the clustering close to $\lambda_c$ seen in Figure 1(b) is an artifact of mechanisms in the particular GA that was used rather than a result of any computational advantage conferred by the $\lambda_c$ regions. We have also performed a wide range of additional experiments to test the robustness of our results. Not only have they held up, but these experiments have pointed to a number of mechanisms that control the interaction of evolution and computation.

**What Causes the Dip at $\lambda = 1/2$?**

Aside from the many differences between Figure 1(b) and Figure 1(c), there is one rough similarity: the histogram shows two symmetrical peaks surrounding a central dip. We found that in our experiment this feature is due to a kind of symmetry breaking on the part of the GA; this symmetry breaking actually impedes the GA's ability to find a rule with performance at the level of the GKL rule. In short, the mechanism is the following. On each run, the best strategy found by the GA is one of two equally fit strategies:

> **Strategy 1**: If the initial configuration contains a sufficiently large block of adjacent (or nearly adjacent) 1's, then increase the size of the block until the entire lattice consists of 1's. Otherwise, quickly relax to a configuration of all 0's.

> **Strategy 2**: If the initial configuration contains a sufficiently large block of adjacent (or nearly adjacent) 0's, then increase the size of the block until the entire lattice consists of 0's. Otherwise, quickly relax to a configuration of all 1's.

These two strategies rely on local inhomogeneities in the initial configuration as indicators of $\rho$. Strategy 1 assumes that if there is a sufficiently large block of 1's initially, then the $\rho$ is likely to be greater than 1/2, and is otherwise likely to be less than 1/2. Strategy 2 makes similar assumptions for sufficiently large blocks of 0's. Such strategies are vulnerable to a number of classification errors. For example, a rule might *create* a sufficiently sized block of 1's that was not present in an initial configuration with $\rho < 1/2$ and increase its size to yield an incorrect final configuration. But, as is explained in [24], rules with $\lambda < 1/2$ (for Strategy 1) and rules with $\lambda > 1/2$ (for Strategy 2) are less vulnerable to such errors than are rules with $\lambda = 1/2$. A rule with $\lambda < 1/2$ maps more than half of the neighborhoods to 0 and thus tends to decrease the initial $\rho$. Due to this it is less likely to *create* a sufficiently sized block of 1's from a low-density initial configuration.

The symmetry breaking involves deciding whether to increase blocks of 1's or blocks of 0's. The GKL rule is perfectly symmetric with respect to the increase of blocks of 1's and 0's. The GA on the other hand tends to discover one or the other strategy, and the one that is discovered first tends to take over the population, moving the population $\lambda$'s to one or the other side of 1/2.

The shape of the histogram in Figure 1(c) thus results from the combination of a number of forces: the selection and combinatorial drift forces described above push the population toward $\lambda = 1/2$, and the error-resisting forces just described push the population away from $\lambda = 1/2$. (Details of the epochs the GA undergoes in developing these strategies are described in [24].)

It is important to understand how in general such symmetry breaking can impede an evolutionary process from finding optimal strategies. This is a subject we are currently investigating.

## 6. Conclusion

In this paper we have reviewed some general ideas about the relationship between dynamical systems theory and the theory of computation. In particular, we have discussed in detail work by Langton and by Packard on the relation between dynamical behavior and computation in cellular automata. Langton investigated correlations between $\lambda$ and CA behavior as measured by several statistics, and Packard's experiment was meant to directly test the hypothesis that computational ability is correlated with $\lambda_c$ regions of CA rule space.

We have presented theoretical arguments and results from an experiment similar to Packard's. From these we conclude that the original interpretation of Packard's results is not correct. We believe that those original results were due to mechanisms in the particular GA used in [29] rather than to intrinsic computational properties of $\lambda_c$ CA.

The results presented here do not disprove the hypothesis that computational capability can be correlated with phase transitions in CA rule space.[4] Indeed, this general phenomena has already been noted for other dynamical systems, as noted in the introduction [10]. More generally, the computational capacity of evolving systems may very well require dynamical properties characteristic of phase transitions if they are to increase their complexity.

---

[4]There are some results concerning computation in CA and phase transitions. Individual CA have been known for some time to exhibit phase transitions with the requisite divergence of correlation length required for infinite memory capacity.[5]

We have shown only that the published experimental support cited for hypotheses relating $\lambda_c$ and computational capability in CA was not reproduced. One problem is that these hypotheses have not been rigorously formulated. If the hypotheses put forth in [22] and [29] are interpreted to mean that *any* rule performing complex computation (as exemplified by the $\rho = 1/2$ task) must be close to $\lambda_c$, then we have shown it to be false with our argument that correct performance on the $\rho = 1/2$ task requires $\lambda = 1/2$. If, instead, the hypotheses are concerned with generic, statistical properties of CA rule space—the "average" behavior of an "average" CA at a given $\lambda$—then the notion of "average behavior" must be better defined. Additionally more appropriate measures of dynamical behavior and computational capability must be formulated, and the notion of the "edge of chaos" must also be well-defined.

Static parameters estimated directly from the equations of motion, as $\lambda$ is from the CA rule table, are only the simplest first step at making such hypotheses and terms well-defined. $\lambda$ and $\gamma$ are excellent examples of the problems one encounters: their correlation with dynamical behavior is weak; they have far too much variance when viewed over CA space; and so on. What is needed is a more structural analysis that goes beyond measuring degrees of randomness and that allows one to detect the intrinsic computational capability in CA behavior. This need is all the more salient in light of the analysis given here, shows that there are problems with using any *particular* computational task to test statistical hypotheses relating $\lambda$ to computational ability. Any particular task is likely to require CA with a particular range of $\lambda$ values for good performance, and the particular range required is a function only of the particular task, not of intrinsic properties of regions of CA rule space.

Let us close by re-emphasizing that our studies do not preclude a future rigorous and useful definition of the phrase "edge of chaos" in the context of cellular automata. Nor do they preclude the discovery that it is associated with a CA's increased computational capability. Finally, they do not preclude adaptive systems moving to such dynamical regimes in order to take advantage of the intrinsic computational capability there. In fact, the present work is motivated by our interest in this last possibility. And the immediate result of that interest is this attempt to clarify the underling issues in the hope of facilitating new progress along these lines.

## Acknowlegments

This research was supported by the Santa Fe Institute, under the Core Research, Adaptive Computation and External Faculty Programs, and by the University of California, Berkeley, under contract AFOSR 91-0293. Thanks to Doyne Farmer, Jim Hanson, Erica Jen, Chris Langton, Wentian Li, Cris Moore, and Norman Packard for many helpful discussions and suggestions concerning this project. Thanks also to Emily Dickinson and Terry Jones for technical advice.